\documentclass[%
reprint,
 amsmath,amssymb,
 aps,
 prb,
]{revtex4-1}
\usepackage{graphicx}% Include figure files
\usepackage{subcaption}
\usepackage{dcolumn}% Align table columns on decimal point
\usepackage{bm}% bold math
\usepackage{miller}
\usepackage{epstopdf}
\usepackage{mathrsfs}
\usepackage{mhchem}
\graphicspath{ {Figures/} }
	
\usepackage{color}
\def\etal{\emph{et al}.}

\newcommand*{\rtten}[1]{\mathbf{\boldsymbol{#1}}}
\newcommand*{\rtvec}[1]{\mathbf{#1}}

\begin{document}

\preprint{APS/123-QED}

\title{Role of Anisotropy in Determining Stability of Electrodeposition at Solid-Solid Interfaces}

\author{Zeeshan Ahmad}
\affiliation{%
 Department of Mechanical Engineering, Carnegie Mellon University, Pittsburgh, Pennsylvania 15213, USA
}
\author{Venkatasubramanian Viswanathan}
 \email{venkvis@cmu.edu}
\affiliation{%
 Department of Mechanical Engineering, Carnegie Mellon University, Pittsburgh, Pennsylvania 15213, USA}
 \affiliation{%
 Department of Physics, Carnegie Mellon University, Pittsburgh, Pennsylvania 15213, USA}

\date{\today}

\begin{abstract}
We investigate the stability of electrodeposition at solid-solid interfaces for materials exhibiting an anisotropic mechanical response. The stability of electrodeposition or resistance to formation of dendrites is studied within a linear stability analysis. The deformation and stress equations are solved using the Stroh formalism and faithfully recover the boundary conditions at the interface. The stability parameter is used to quantify the stability of different solid-solid interfaces incorporating the full anisotropy of the elastic tensor of the two materials. Results show a high degree of variability in the stability parameter depending on the crystallographic orientation of the solids in contact and points to opportunities for exploiting this effect in developing Li metal anodes.
\end{abstract}

\maketitle

\section{Introduction}\label{sec:intro}
Solid-solid interfaces are ubiquitous in several important engineering applications like epitaxial thin films~\cite{freund2004thin}, solid-state batteries \cite{Kato2016-ssb, solid-polymer-electrolytes-2008}, solid-oxide fuel cells \cite{ormerod2003solid}, solid-state electrolysis \cite{jordan1973purification, Lambert2002solid} and are observed in nature in sedimentary rocks and porous materials with the formation of irregular interfaces called stylolites~\cite{Angheluta2008}. In particular, electrodeposition at solid-solid interfaces is of great interest due to the possibility of obtaining safer and higher energy density batteries based on Li and other metal anodes. These anodes rely on plating rather than intercalation and  have eluded all attempts at functioning in liquid-solid interfaces due to unstable surface layer and electrodeposition resulting in formation of dendrites \cite{aurbach2000, aurbach-2002-failure, balsara2014-subsurface, PhysRevLett.56.1260, steiger2014dendrite}. Dendrites have also been observed during electrodeposition at solid-solid interfaces like during solid-state electrolysis of Cu \cite{Lambert2002solid} and Li-garnet solid electrolyte interface \cite{Ren2015direct}.  Controlling the growth of dendrites during electrodeposition at a solid electrolyte-metal interface could enable the use of metal anodes, especially Li \cite{XuLi2014}, on which several high energy density batteries like Li-\ce{O2} and Li-S rely\cite{bruceLiOS2012, christensen2011-Liair, oleg2015TMR}. Solid electrolyte systems also offer the advantages of improved safety, absence of leakage, better chemical and mechanical stability, and possibility of miniaturisation through thin film processing techniques for integration in devices like laptops and cellphones \cite{wang2015design, knauth2002solid}. 
%Typically, reactions and phase transformations occur at the interface and the ingredients of a continuum mechanics formulation of phase transformation at solid-solid interfaces are: kinematics, mechanics, thermomechanics, constitutive theory and kinetics~\cite{Gurtin1993}. 

The stresses generated at the interface between the solids are expected to play a major role in the reactions occurring at solid-solid interfaces. The key role of interfacial stresses in affecting the rates of electrodeposition was analyzed in a seminal work by Monroe and Newman \cite{Monroe2004Effect}.  They further analyzed the interfacial stability of Li-solid polymer electrolyte system within linear elasticity theory and showed using a kinetic model that solid polymer electrolytes with a sufficient modulus are capable of suppressing dendrite growth\cite{Monroe2005Impact}.  In a recent work, we extended the analysis to include the case of inorganic solid electrolytes \cite{ahmad2017stability}.  The key difference between a solid polymer electrolyte and a solid inorganic electrolyte is that they possess vastly different partial molar volume of Li, which strongly affects the role of hydrostatic stresses at the interface \cite{ahmad2017stability}.   We showed the existence of a new stable regime that is a density-driven stabilizing mechanism in addition to the pressure-driven stability mechanism identified earlier.  However, both of these earlier analyses invoke the assumption of isotropic elastic response for the two solid materials. This assumption may generally not hold for the metal phase, Li since it has an anisotropy factor of 8.52~\cite{xuLi2017}, and the solid electrolyte phase \cite{Ahmad16uncertainty}. The shear and elastic modulus vary by a factor of $\sim$4 between the stiffest and most compliant directions. The anisotropy arises when the interface properties are dominated by one particular crystal orientation rather than an average over all crystal orientations. This may occur, for example, when the surface of the solid in contact is single crystalline. Even for bulk isotropic materials, the local mechanical response may be anisotropic dictated by the crystallographic orientation of the surfaces in contact \cite{freund2004thin}. In such a case, the anisotropic stress-strain relations corresponding to the crystal orientation should be used \cite{ting-anis, bower2009mechanics}. In this work, we relax the assumption of isotropy and analyze electrodeposition at solid-solid interfaces for anisotropic elastic materials.

In Ref.~\onlinecite{ahmad2017stability}, we have constructed a generalized two parameter stability diagram of electrodeposition for isotropic solid-solid interfaces. In this work, along similar lines, we develop a continuum mechanics-based theory for the analyzing stability of electrodeposition at interfaces with anisotropic mechanical response. The interface stability is studied using a linear stability analysis similar to the Mullins-Sekerka \cite{MullinsSekerka63, MullinsSekerka64} and Asaro-Tiller-Grinfeld approach \cite{Asaro1972, grinfeld1986} while incorporating the full anisotropy of the elastic tensor of the materials at the interface. The Stroh formalism used faithfully captures the boundary conditions of perturbation imposed in the linear stability analysis as well as vanishing of deformations and stresses far from the interface. The differences between the three cases of isotropic-isotropic, isotropic-anisotropic and fully anisotropic interface are highlighted through the deformation fields obtained and the stability diagrams. As expected, the partial molar volume/density of the metal in the solid electrolyte greatly changes the condition for stable electrodeposition \cite{Angheluta2009, ahmad2017stability}. This paper is organized as follows. In Sec. \ref{sec:theory}, we develop the treatment of anisotropy in the elastic tensor used throughout the paper. Some differences encountered in electrodeposition at solid-solid interfaces are also highlighted. In Sec. \ref{sec:lsa}, we apply the Stroh formalism to solve deformation equations obtained on a linear perturbation. In Sec. \ref{sec:results}, we generate stability diagrams and calculate the stability parameter which is a measure of stability. In Sec. \ref{sec:disc}, we comment and discuss some general principles of stability diagrams obtained. We end with concluding remarks in Sec. \ref{sec:concl}.

\section{Theory}\label{sec:theory}

In this section, we develop the procedure used to compute the deformation and stress profiles for anisotropic materials including the crystallographic orientation dependent elastic matrix computations, and theory of electrodeposition at solid-solid interfaces.

\subsection{Stroh Formalism}
The Stroh formalism \cite{stroh1958, stroh1962}, based on the Eshelby-Read-Shockley formalism \cite{erd1953} is a mathematically powerful tool for solving two-dimensional problems in anisotropic linear elasticity. A two-dimensional analysis should suffice for our problem of determining stability, since the major features required for determining stability can be generated - perturbation of a given wavenumber, surface tension, interfacial stresses etc. In what follows, we shall develop the Stroh formalism for a two-dimensional elasticity problem \cite{ting-anis} and explicitly write down the expressions for the deformation and stress fields in terms of the elastic tensor of the material.

We denote the deformation and stress fields by $\rtvec{u}$ and $\rtten{\sigma}$. For force balance, the necessary condition for the stress field is:
\begin{equation}\label{eq:forbal}
\text{div}(\rtten{\sigma})=\rtvec{0}.
\end{equation}
The stress can be related to the deformation field using the linear elasticity relationship for anisotropic materials:
\begin{eqnarray}\label{eq:const}
\sigma_{ij}=C_{ijkl}u_{k,l}.
\end{eqnarray}
Here $u_{k,l}$ indicates differentiation of $u_k$ with respect to $x_l$ i.e. $u_{k,l}=\partial u_k/\partial x_l$. Subsequently, we use , to indicate differentiation with respect to indices placed after it. The repeated indices are summed over as in the Einstein summation convention. Substituting the stress from Eq. (\ref{eq:const}) into the force balance and using the symmetries of $\rtten{\sigma}$, we obtain:
\begin{eqnarray}\label{eq:forbal1}
C_{ijkl}u_{k,lj}=0.
\end{eqnarray}

For the two-dimensional problem, Eq. (\ref{eq:forbal1}) is a second order homogeneous differential equation in the independent variable $x_1$ and $x_2$. The deformation $\rtvec{u}$ will generically depend on a linear combination of $x_1$ and $x_2$ i.e. $\rtvec{u}=\rtvec{a} f(x_1+px_2)$. Differentiating $u_k$ with respect to $x_l$ and $x_j$, and plugging in Eq. (\ref{eq:forbal1}), we get
\begin{eqnarray}
C_{ijkl}(\delta_{j1}+p\delta_{j2})(\delta_{l1}+p\delta_{l2})a_k=0\\
\implies \left(C_{i1k1}+p(C_{i1k2}+C_{i2k1})+p^2C_{i2k2}\right)a_k=0\label{eq:Cforbal}
%\implies \left( Q+p(R+R^T)+p^2T \right)=0
\end{eqnarray} 
In terms of the tensors $R_{ik}=C_{i1k1}$, $S_{ik}=C_{i1k2}$ and $T_{ik}=C_{i2k2}$, Eq. (\ref{eq:Cforbal}) becomes
\begin{equation}\label{eq:Qforbal}
\implies \left( \rtten{R}+p(\rtten{S}+\rtten{S}^T)+p^2\rtten{T} \right)\rtvec{a}=\rtvec{0}.
\end{equation}
This is an eigen value equation with eigen value zero and eigen vector $\rtvec{a}$. For solutions to exist, we must have:
\begin{eqnarray}\label{eq:degree6}
\det(\rtten{R}+p(\rtten{S}+\rtten{S}^T)+p^2\rtten{T} )=0.
\end{eqnarray}
This gives a sixth degree equation which can be solved for $p$. The stress tensor associated with this deformation can be calculated using
\begin{subequations}
\begin{eqnarray}
\sigma_{i1}=(R_{ik}+pS_{ik})a_kf'(x_1+px_2)\\
\sigma_{i2}=(S_{ki}+pT_{ik})a_kf'(x_1+px_2)
\end{eqnarray}
\end{subequations}
The stress can be written in terms of the stress function $\varphi$:
\begin{eqnarray*}
\varphi_i=b_i f(x_1+p_ix_2); \rtvec{b}=(\rtten{S}^T+p\rtten{T})\rtvec{a}=-\frac{1}{p}(\rtten{R}+p\rtten{S})\rtvec{a}\\
\end{eqnarray*}
\begin{equation}
\sigma_{i1}=-\varphi_{i,2}, \sigma_{i2}=\varphi_{i,1}
\end{equation}

The solutions to $p$ will be complex with a non-zero imaginary part. Since the solutions will occur as complex conjugates, in the absence of degeneracies, we can write the deformation and stress as linear combinations of the individual solutions with Im$(p_\alpha)>0$:
\begin{subequations}\label{eq:ustrresult}
\begin{eqnarray}
\rtvec{u}=2\text{Re}\left\{\sum_{\alpha=1}^3 q_{\alpha} \rtvec{a_{\alpha}} f_{\alpha}(x_1+p_{\alpha} x_2)\right\}\\
\rtvec{\varphi}=2\text{Re}\left\{\sum_{\alpha=1}^3 q_{\alpha} \rtvec{b_{\alpha}} f_{\alpha}(x_1+p_{\alpha} x_2)\right\}
\end{eqnarray}
\end{subequations}
The above result may be written in compact form using the matrices $\rtten{A}=[\rtvec{a_1}\  \rtvec{a_2}\  \rtvec{a_3}]$,  $\rtten{B}=[\rtvec{b_1}\  \rtvec{b_2}\  \rtvec{b_3}]$, $\rtten{F}=\text{diag}[f(x_1+p_1x_2)\ f(x_1+p_2x_2)\ f(x_1+p_3x_2)]$ and constants $\rtvec{q}=[q_1\ q_2\ q_3]^T$:
\begin{subequations}\label{eq:ustresultmat}
\begin{eqnarray}
\rtvec{u}=2\text{Re}\left\{ \rtten{A}\rtten{F}\rtvec{q} \right\}\\
\rtvec{\varphi}=2\text{Re}\left\{ \rtten{B}\rtten{F}\rtvec{q} \right\}
\end{eqnarray}
\end{subequations}

The procedure for degenerate case of isotropic material is mentioned in appendix \ref{app:degen}.

\subsection{Electrodeposition at solid-solid interfaces}
During electrodeposition, the metal ions present in the solid electrolyte are reduced at the metal anode according to the reaction:
\begin{eqnarray}\label{eq:reaction}
\mathrm{{M}^{z+}+ze^{-}}\rightleftharpoons\mathrm{M}.
\end{eqnarray}

\begin{figure}[htbp]
{\includegraphics[width=0.45\textwidth]{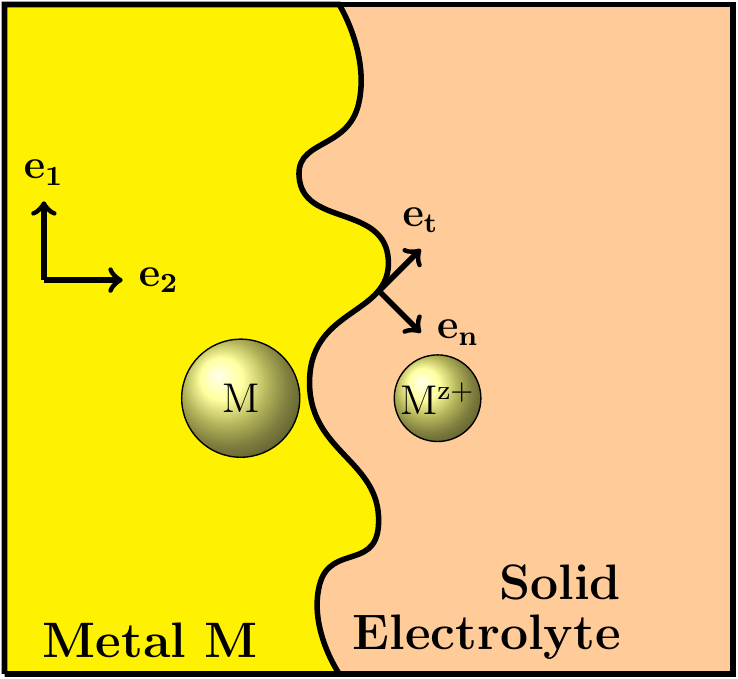}}
\caption{\label{fig:schematic} (color online). Schematic of the electrodeposition problem with metal electrode-solid electrolyte interface. The metal surface $x_2=f(x_1,t)$ grows on deposition of metal ions, the rate of which is proportional to the current. The local geometry alters the kinetics of deposition at the interface.}
\end{figure}

The metal surface $x_2=f(x_1,t)$ grows in response to current density $i$ normal to the metal surface (Fig .\ref{fig:schematic}). The current density without any deformation can be related to the surface overpotential $\eta$ through the Butler-Volmer equation \cite{newman2012electrochemical}
\begin{eqnarray}\label{eq:i}
\frac{i}{i_0}= \left[ \exp \left(\frac{\alpha_a \mathrm{z}F \eta}{RT} \right)  - \exp \left(-\frac{\alpha_c \mathrm{z}F \eta}{RT} \right)  \right].
\end{eqnarray}
Here $\alpha_a$ and $\alpha_c$ are the charge transfer coefficients associated with anodic and cathodic reactions, and $i_0$ is the exchange current density. The current density at a deformed interface can be written in terms of the undeformed current density as:
\begin{eqnarray}\label{eq:inew}
\frac{i_{\text{deformed}}}{i_{\text{undeformed}}}=\exp\left[\frac{(1-\alpha_a)\Delta \mu_{e^-}}{RT} \right]. 
\end{eqnarray}
where $\Delta \mu_{e^-}$ is the change in electrochemical potential of the electron due to deformation at the interface given by\cite{Monroe2004Effect}:
\begin{eqnarray}\label{eq:mu}
\begin{split}
\Delta \mu_{e^-}=&-\frac{V_\mathrm{{M}}}{2\mathrm{z}}\left(1 + v \right) \left(-\gamma \kappa \right.  \\
&\left. - \rtvec{e_n}\cdot [(\rtten{\tau_e} - \rtten{\tau_s}) \rtvec{e_n}]\right)\\
& + \frac{V_\mathrm{{M}}}{2\mathrm{z}} \left(1 - v \right) \left(\Delta p_e + \Delta p_s \right).
\end{split}
\end{eqnarray}
Here $V_{\mathrm{M}}$ is the molar volume of metal species in metallic form, $v=V_\mathrm{{M^{z+}}}/V_\mathrm{{M}}$ is ratio of molar volume of the metal ion in the solid electrolyte to that in the metal, $\gamma$ is the surface tension, $\kappa$ is the mean curvature at the interface, $\rtten{\tau_e}$ and $\rtten{\tau_s} $ are the deviatoric stresses at the electrode and electrolyte sides of the interface,  and $\Delta p_e$ and $\Delta p_s$ are the gage pressures at the electrode and electrolyte sides of the interface at $x_2=0$. Hereafter, we use the subscripts $s$ and $e$ for properties of the solid electrolyte and electrode respectively. Eq. (\ref{eq:mu}) is obtained by calculating the electrochemical potential change $d\mu=(\partial \mu/\partial p) dp$ and using the equilibrium of Eq. (\ref{eq:reaction})\cite{Monroe2004Effect}. While performing linear stability analysis, we retain terms only up to first order in perturbation, which removes the second order strain energy density terms commonly encountered in the Asaro-Tiller-Grinfield formalism \cite{Asaro1972, grinfeld1986}. The first term is due to surface tension, while the second and third terms are due to deviatoric and hydrostatic stresses at the interface. The negative sign in the deviatoric term is different from the positive sign obtained by Monroe and Newman \cite{Monroe2004Effect} since we use the convention of decomposition of stress as $-p\rtten{I}+\rtten{\tau}$, rather than $p\rtten{I}+\rtten{\tau}$ used by them. 

The surface tension provides a stabilizing mechanism against roughening of the interface. It increases the electrochemical potential at peaks in the propagating interface and decreases it in the valleys. However, at length scales of roughening encountered in electrodeposition problems, the stabilization by surface tension is much smaller compared to that by interfacial stresses \footnote{See for example, Fig. 5 in Ref \onlinecite{Monroe2005Impact} or the expression for contributions to the stability parameter in Supplemental Material of Ref. \onlinecite{ahmad2017stability}}. For this reason, we have ignored the contribution of surface tension term in the electrochemical potential throughout this paper. The electrochemical potential is determined by an interplay between the deviatoric and hydrostatic terms.

Electrodeposition at solid-solid interfaces has several advantages compared to that at liquid-solid interfaces, especially for applications in batteries. Besides the stabilization of the propagating interface by interfacial stresses, solid electrolytes have a cation transference number close to 1. The mechanism of dendritic growth due to depletion of ions, common in liquid electrolytes, is thus not an issue when solid electrolytes are used.
The Sand's time, which is a measure of the time it takes for ion depletion at the negative electrode, approaches infinity, since it is proportional to the transference number of the anion \cite{monroe2003-dendrite}. The depletion of ions is a major cause of dendrite growth. The growth rate of dendrites, which is proportional to the mobility of the anion, becomes zero \cite{Chazalviel90}. 
%Another advantage of electrodeposition at solid-solid interfaces involving solid polymer electrolytes is that the unlike liquid electrolytes, they have don't form solid electrolyte interphase (SEI) which might be detrimental for the Coulombic efficiency of the electrochemical process \cite{XuLi2014}. 
%Based on calculations for a solid polymer with $v>1$, Monroe and Newman predicted that a solid polymer electrolyte (SPE) with a high shear modulus can suppress uneven electrodeposition. This SPEs are stable against metal anodes like Li and, therefore, a mechanically strong SPE in contact with a metal anode can be used to suppress dendrite growth.

\subsection{Transformation of the elastic tensor}\label{sec:ctrans}
\begin{figure}[htbp]
{\includegraphics[width=0.45\textwidth]{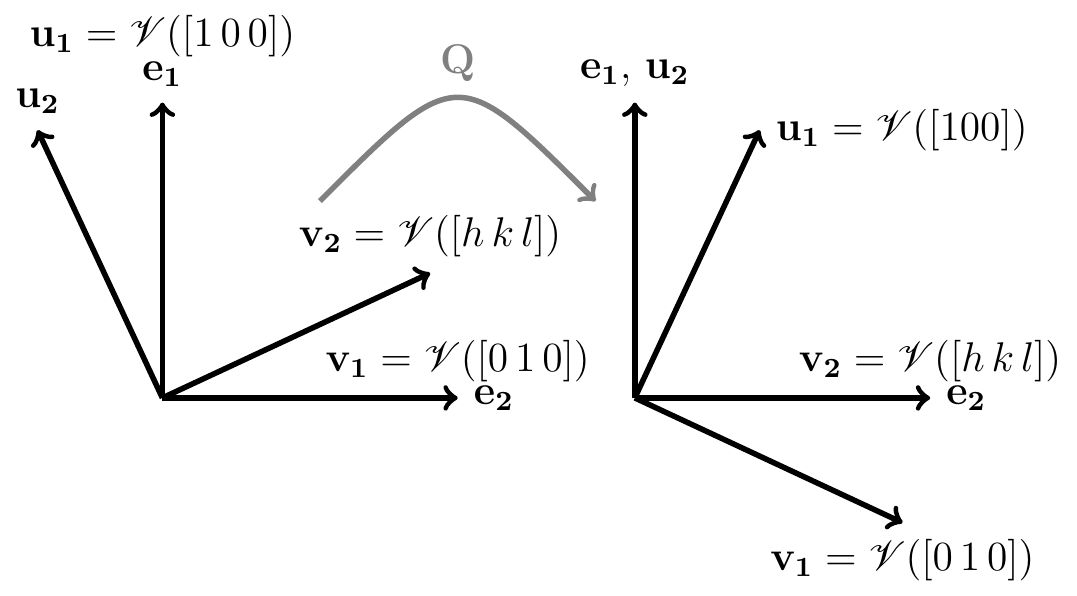}}
\caption{\label{fig:Qtrans} Transformation of the elastic tensor for anisotropic analysis. The rotation brings the required crystallographic directions $\rtvec{v_2}$ and $\rtvec{u_2}$ along $\rtvec{e_2}$ and $\rtvec{e_1}$. Note that the rotation operation is performed on the crystallographic axes of the material and not on the actual axes considered in the problem $\rtvec{e_1}$ and $\rtvec{e_2}$.}
\end{figure}

For anisotropic analysis, the crystallographic directions of the electrode and electrolyte along $\rtvec{e_1}$ and $\rtvec{e_2}$ will determine the mechanical response through the elastic tensor. When the surfaces of the electrode and electrolyte in contact are such that the crystallographic axes of the material coincide with the actual axes (in Fig. \ref{fig:schematic}), the elastic tensor can be plugged into the equations directly. This is the case when $\hkl[0 1 0]$ and $\hkl[1 0 0]$ crystallographic directions of the material are aligned along $\rtvec{e_2}$ and $\rtvec{e_1}$ respectively. If some other crystallographic directions are along $\rtten{e_1}$ and $\rtten{e_2}$, the elastic tensor has to be transformed according to the rotation matrix $\rtten{Q}$ that aligns the required crystallographic directions along $\rtvec{e_1}$ and $\rtvec{e_2}$ as shown in Fig. \ref{fig:Qtrans}. Once the rotation matrix $\rtten{Q}$ is obtained, the elastic tensor can be transformed according to:
\begin{eqnarray}\label{eq:ctransf}
\tilde{C}_{ijkl}=Q_{ip}Q_{jq}Q_{kr}Q_{ls}C_{pqrs}
\end{eqnarray}
An analogue of Eq. (\ref{eq:ctransf}), given in appendix \ref{app:ctrans}, can be used to transform the elastic tensor in Voigt form as well.

\emph{Determination of $\rtten{Q}$}.--We have seen that the problem of determination of the elastic tensor reduces to determination of the rotation matrix $\rtten{Q}$. Let $\mathscr{V}$ (\hkl[h k l]) denote the direction vector corresponding to crystallographic direction \hkl[h k l]. For example, $\mathscr{V}(\hkl[1 0 0])=(a,0,0)$ for a cubic crystal and $\mathscr{V}(\hkl[1 1 1])=(a,b,c)$ for an orthorhombic crystal where $a$, $b$ and $c$ are the respective lattice constants. For our calculations, we treated the crystallographic direction of the material along $\rtvec{e_2}$, referred to as $\rtvec{v_2}$, as the independent direction. Then, $\rtten{Q}$ is obtained as the rotation that aligns $\rtvec{v_2}$ along $\rtvec{v_1}=\mathscr{V}(\hkl[0 1 0])$ (Fig. \ref{fig:Qtrans}). The transformation $\rtten{Q}$ is unique since it is a right-handed rotation about axis $\rtvec{v_2}\times \rtvec{v_1}$ that transforms $\rtvec{v_2}$ to $\rtvec{v_1}$. The new crystallographic direction along $\rtvec{e_1}$, referred to as $\rtvec{u_2}$ is the dependent direction and can be obtained using: $\rtvec{u_2}=\rtten{Q}^{-1}\rtvec{u_1}$ where $\rtvec{u_1}=\mathscr{V}(\hkl[1 0 0])$. An example for a cubic crystal is shown in appendix \ref{app:ctrans}. Finally, we note that the elastic tensor depends not only on the crystallographic direction perpendicular to the interface (i.e. $\rtten{v_2}$) but also on the crystallographic direction along $\rtvec{e_1}$ (i.e. $\rtten{u_2}$).

\section{Linear Stability Analysis}\label{sec:lsa}
A linear stability analysis can be used to determine the growth of various Fourier components of an arbitrary perturbation of the interface. It provides the boundary conditions to the two-dimensional deformation problem. Initially, the electrode is located at $x_2<0$ and the electrolyte at $x_2>0$. The solids are assumed to be in contact at all times i.e. $\rtvec{u}_e(x_1,0)=\rtvec{u}_s(x_1,0)$. Additionally, all deformations are assumed to vanish far from the interface i.e. $\lim_{x_2\to \pm \infty}\rtvec{u}(x_1,x_2)=\rtvec{0}$. The interface $x_2=0$ between the solids is perturbed in a sinusoidal fashion:
\begin{equation}
(x_1,0)\mapsto (x_1, A\cos(kx_1))
\end{equation}
where $k$, the wave number of perturbation, was chosen as $10^8$/m \cite{Monroe2005Impact, ahmad2017stability}. In three dimensions, the interface will have two wavenumbers along different directions. To obtain a sinusoidal perturbation of the interface, we try the following ansatz for the functions $f_\alpha$ in Eq. (\ref{eq:ustrresult}):
\begin{eqnarray}\label{eq:flin}
f_{\alpha}(x_1+p_\alpha x_2)=\begin{cases} e^{ik(x_1+p_\alpha x_2)} &  x_2
>0\\
 e^{-ik(x_1+p_\alpha x_2)} &  x_2
<0
\end{cases}\end{eqnarray}
Here $p_\alpha$ are solutions of the sixth degree equation obtained from Eq. (\ref{eq:degree6}). Since $p_\alpha$ are imaginary, the term $e^{\pm |p_\alpha|x_2}$ represents the decay of the perturbation as we move away from the interface. A straightforward calculation of $\rtvec{u}$ shows that the deformation obtained using this ansatz gives the required perturbation, while also vanishing far from the interface. A tangential force balance at the interface is also imposed:
\begin{equation}
\rtvec{e_t} \cdot \rtten{\tau}_{e}\rtvec{e_n} = \rtvec{e_t}\cdot \rtten{\tau}_{s}\rtvec{e_n}.
\end{equation}

To assess the stability of electrodeposition, we solved for the deformations and stresses using the Stroh formalism and the elastic tensor of the material. Once these quantities are known, one can compute $\Delta \mu_{e^-}$ using Eq. (\ref{eq:mu}). From linear stability analysis, $\Delta \mu_{e^-}(x_1)$ has a form similar to the perturbation i.e.  $\Delta \mu_{e^-}(x_1)=A\chi \cos(kx_1)$. From Eq. (\ref{eq:inew}), we see that the current will promote roughening when $\chi>0$ and reduce roughening when $\chi<0$. This analysis is quite similar to that of Asaro and Tiller for instability during stress corrosion cracking with different kinetics of growth \cite{Asaro1972}. A similar result exists for the stability of stressed interface \cite{freund2004thin}. The interface is stable if the chemical potential of the material increases in the direction of growth. The condition $\chi<0$ ensures that the chemical potential of the electron at peaks is lower so that the mass transfer of Li is lower at the peaks.

%NOTE:
%$u_3(x,0)=0$ for electrode and electrolyte
%Implemented as $u_s^e(x,0)=u_3^s(x,0)$ and $u_3^e(x,0)=0$\\

Fig. \ref{fig:deform} shows the deformation profiles obtained for three different classes of electrode-electrolyte interfaces: both isotropic, electrolyte isotropic and electrode anisotropic, and both anisotropic on the application of a sinusoidal perturbation at the interface. We observe qualitative differences in the deformation profiles for the three cases. The rate of decay of deformation, which depends on the solutions $p_\alpha$ through Eq. (\ref{eq:flin}), are different due to the different mechanical responses. This can be seen from the  difference in deformations at the boundaries (i.e. as we move away from interface $x_2=0$) of the three cases in Fig. \ref{fig:deform}. For example, the deformation $u_2$ along $x_2$ at $x_2=-20$ nm is positive for (b) but negative for (c) and (d).

\iffalse
\begin{figure*}[htbp]
\begin{subfigure}{0.45\textwidth}
{\includegraphics[width=1\textwidth]{nodisp.pdf}}
\caption{without displacement}
\end{subfigure}
\begin{subfigure}{0.45\textwidth}
{\includegraphics[width=1\textwidth]{iso-iso.pdf}}
\caption{isotropic-isotropic interface}
\end{subfigure}
\begin{subfigure}{0.45\textwidth}
{\includegraphics[width=1\textwidth]{iso-aniso.pdf}}
\caption{isotropic-anisotropic interface}
\end{subfigure}
\begin{subfigure}{0.45\textwidth}
{\includegraphics[width=1\textwidth]{aniso-aniso.pdf}}
\caption{anisotropic-anisotropic interface}
\end{subfigure}
\caption{\label{fig:deform} (color online). Deformation field for different interfaces: isotropic-isotropic, isotropic-anisotropic and anisotropic-anisotropic. The interface materials are: Li-LiI with (c) (010) Li surface in contact with isotropic LiI, (d) (010) Li surface with (010) LiI surface with $\rtvec{v_2}=\mathscr{V}(\hkl[010])$, $\rtvec{u_2}=\mathscr{V}(\hkl[100])$ as in Fig. \ref{fig:Qtrans}.}
\end{figure*}
\fi

\begin{figure*}[htbp]
\includegraphics[width=1\textwidth]{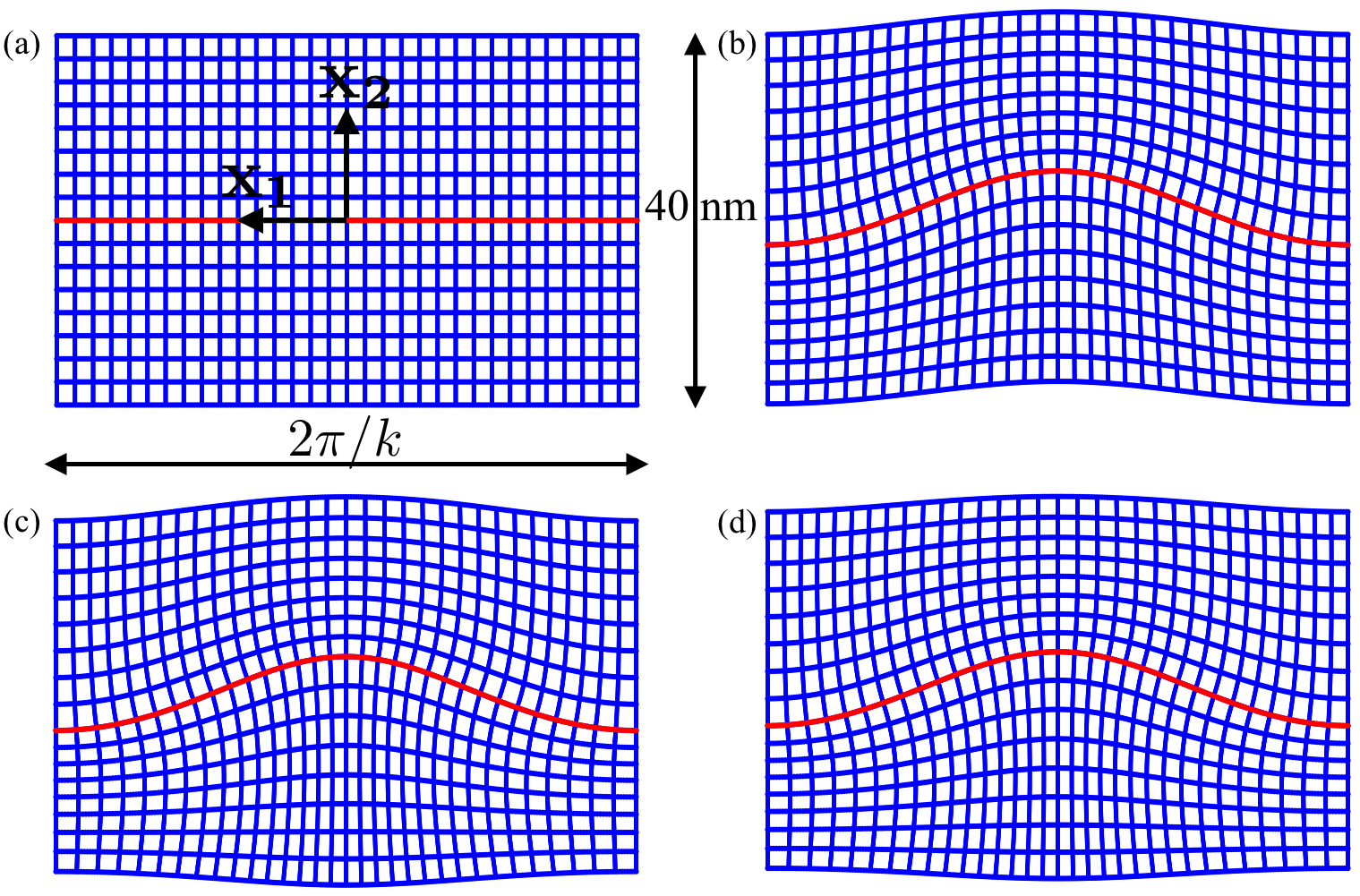}
\caption{\label{fig:deform} Deformation profiles obtained for different mechanical properties at the interface: (a) Initial position and deformed positions for one wavelength $\lambda=2\pi/k=62.8$ nm of interface on application of a sinusoidal perturbation: (b) isotropic-isotropic interface, (c) isotropic-anisotropic interface and (d) anisotropic-anisotropic interface. The interface $x_2=0$ is located in the middle (red line) with electrolyte above ($x_2>0$) and electrode below ($x_2<0$). The materials at the interface are Li electrode and LiI electrolyte. In (c) (010) Li surface is in contact with isotropic LiI, and in (d) (010) Li surface is in contact with (010) LiI surface with $\rtvec{v_2}=\mathscr{V}(\hkl[010])$, $\rtvec{u_2}=\mathscr{V}(\hkl[100])$ as in Fig. \ref{fig:Qtrans}.} The deformations are non-zero at the ends ($x_2=\pm 20$ nm) and vanish only at $x_2=\pm \infty$.
\end{figure*}

\section{Results}\label{sec:results}
We perform calculations of stability parameter for an electrolyte in contact with a Li electrode with shear modulus $G_e=3.4$ GPa, $\nu_e=0.42$, $V_M=1.3\times 10^{-5}$ $\mathrm{m^3}$/mol. For anisotropic Li electrode, the elastic tensor was obtained from CRC handbook \cite{crcelastic}. The extension of results to other electrode materials is straightforward.

\subsection{Isotropic-isotropic interface}
In this case, the stability parameter can be expressed analytically in terms of the two elastic constants - shear modulus $G$ and Poisson's ratio $\nu$ of the isotropic electrode and electrolyte and parameters ($V_\mathrm{M}$, $v$, $\mathrm{z}$) related to the properties of the electrodeposition reaction (\ref{eq:reaction}). The stability diagram has four regions out of which two are stable \cite{ahmad2017stability}. These are high density low shear modulus region, whose stability is density-driven and low density high shear modulus region, whose stability is pressure driven. The details can be found in Ref. \onlinecite{ahmad2017stability}. The stability of low density high shear modulus region was first predicted by Monroe and Newman \cite{Monroe2005Impact} and later validated experimentally by Balsara and co-workers \cite{balsara2012-modadh, balsara2016-mod} in a qualitative sense. In our study, we solved the deformation equations using the Stroh formalism for the degenerate case of isotropic material as shown is Appendix \ref{app:degen}. The results obtained for the stability parameter were the same as Ref.~\onlinecite{Monroe2005Impact} and ~\onlinecite{ahmad2017stability}, thus validating our use of the machinery of Stroh formalism.

\subsection{Isotropic-Anisotropic interface}

\begin{figure}[htbp]
\includegraphics[width=0.45\textwidth]{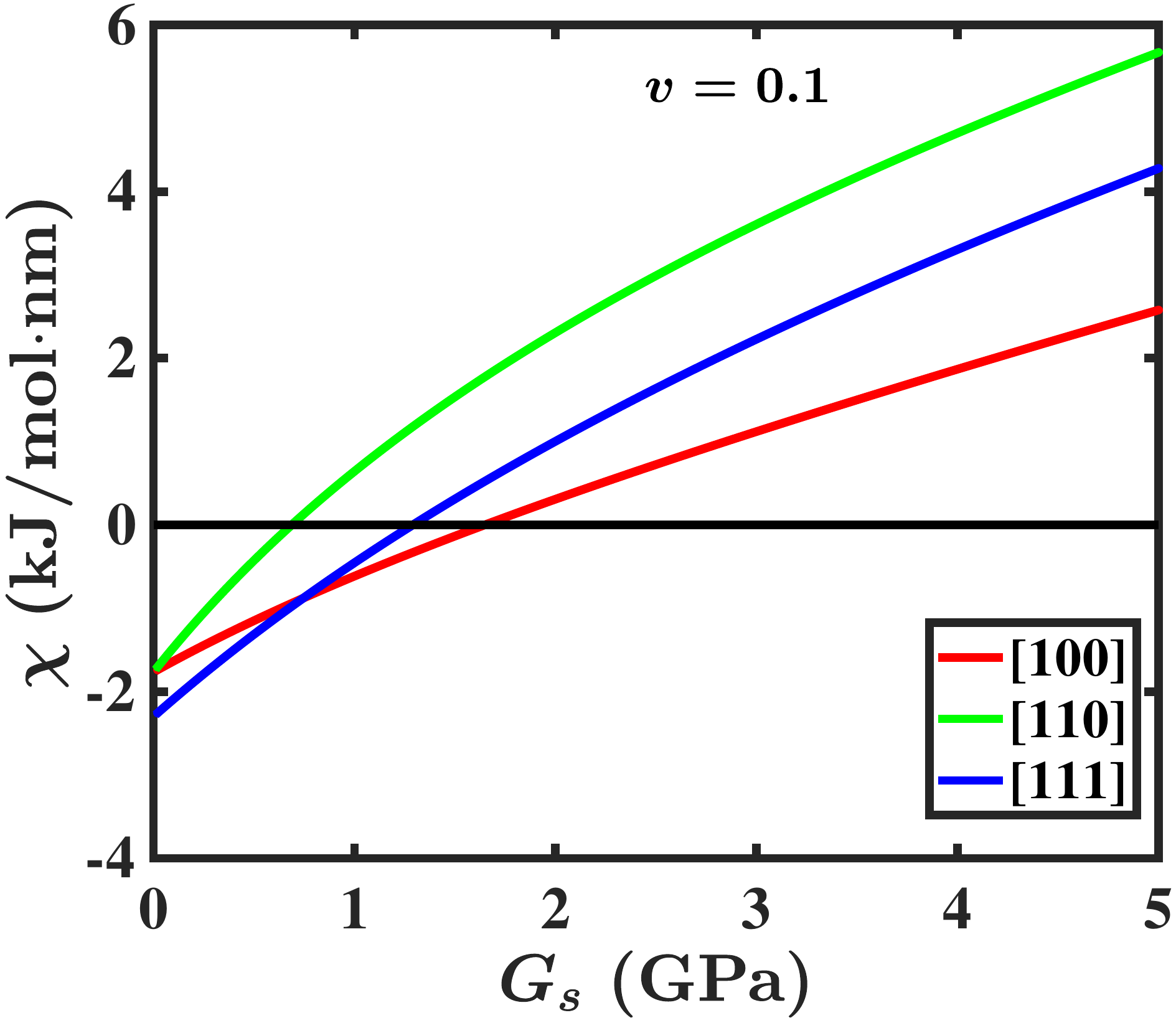}
\includegraphics[width=0.45\textwidth]{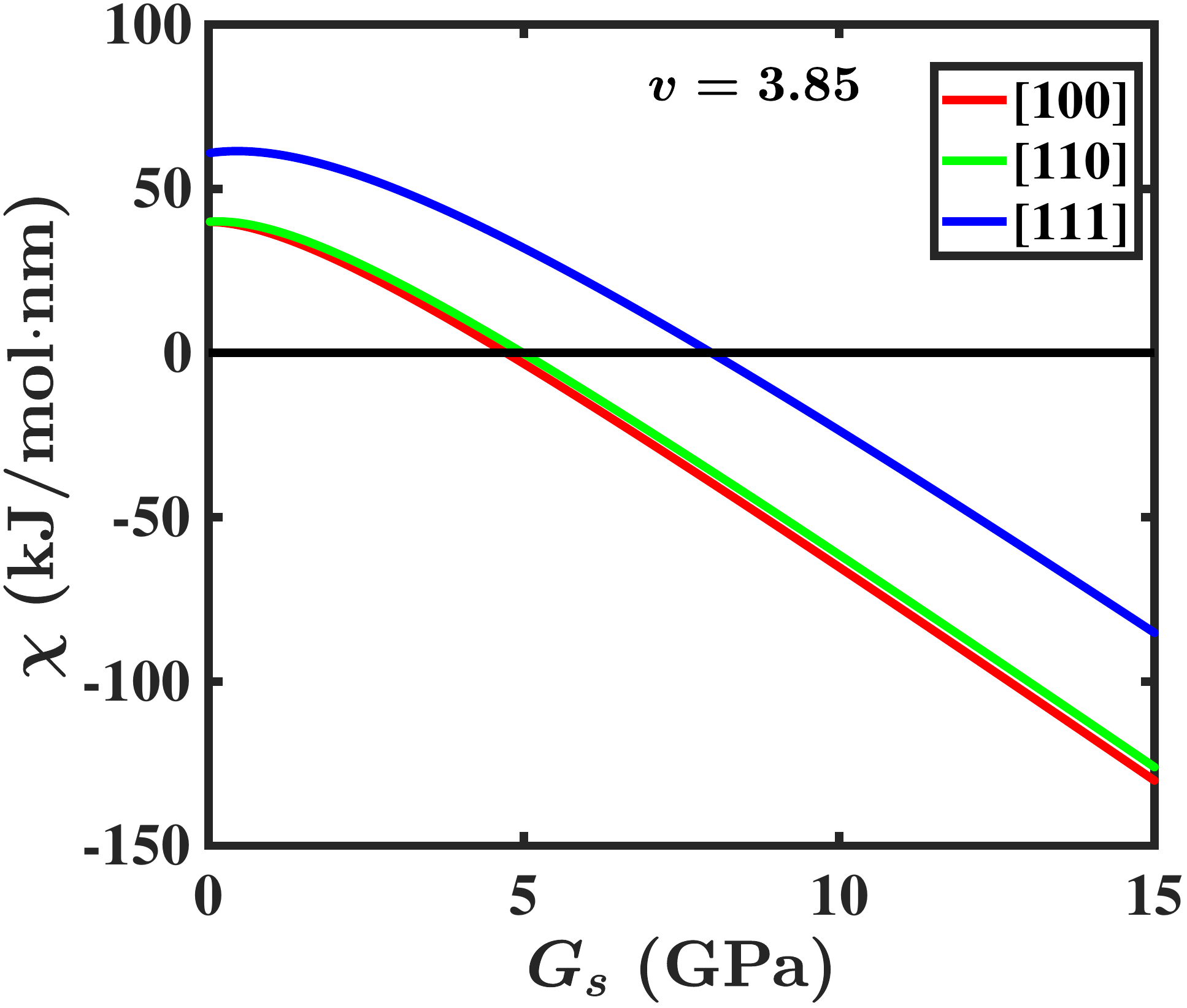}
\caption{\label{fig:chi} (color online). Stability parameter of a solid electrolyte-Li electrode system as a function of the shear modulus of solid electrolyte for $\rtvec{v_2}=\mathscr{V}(\hkl[0 1 0])$, $\mathscr{V}(\hkl[1 1 0])$ and $\mathscr{V}(\hkl[1 1 1])$ and $v=0.1$, $3.85$ respectively.}
\end{figure}

\begin{figure}
%{\includegraphics[width=0.45\textwidth]{../../shear-modulus-window/stab-diag-iso-aniso.pdf}}
{\includegraphics[width=0.45\textwidth]{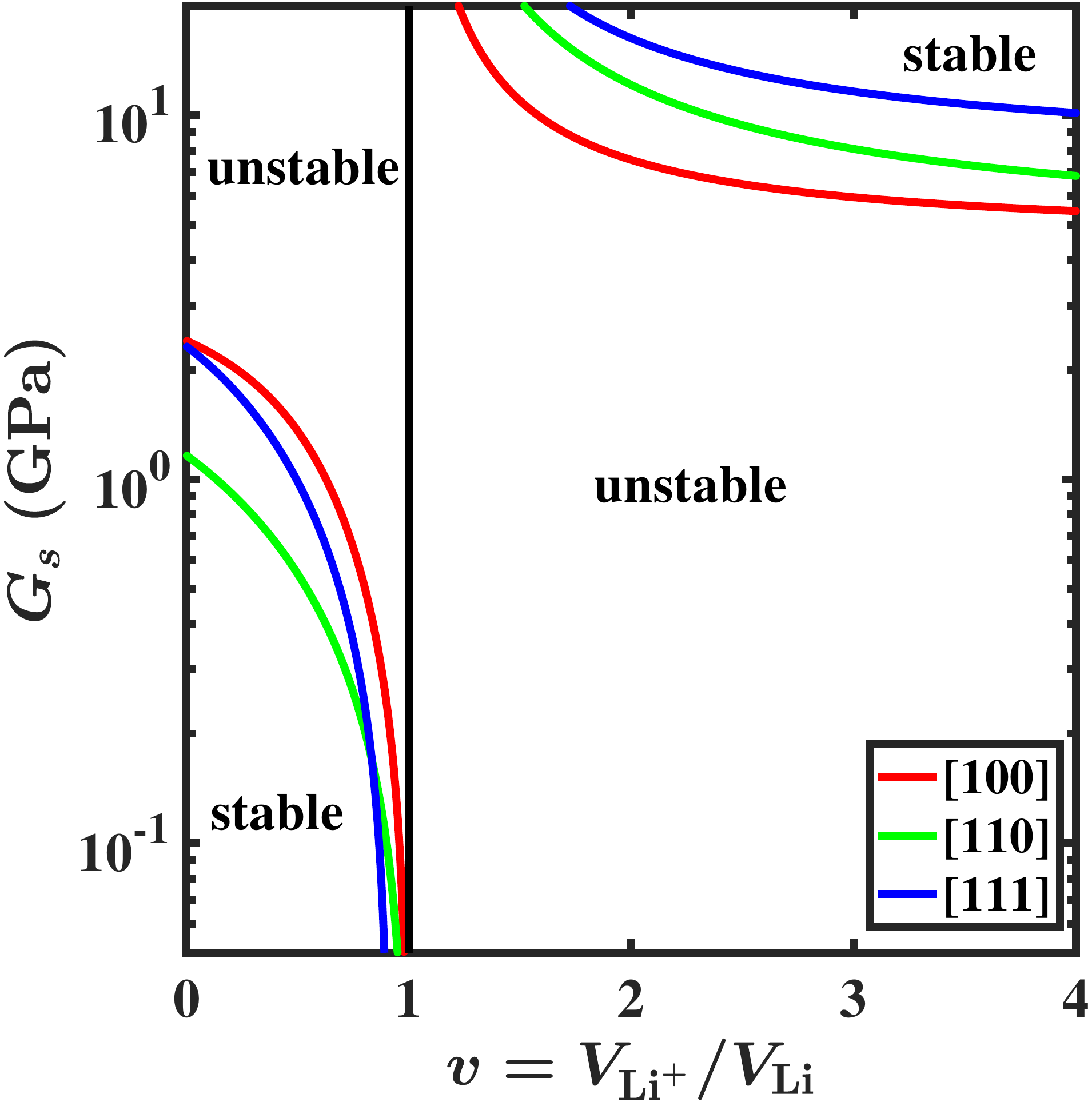}}
\caption{\label{fig:stab-diag-anis} (color online). Stability diagram of isotropic solid electrolyte-anisotropic Li electrode system for $\rtvec{v_2}=\mathscr{V}(\hkl[100])$, $\mathscr{V}(\hkl[110])$ and $\mathscr{V}(\hkl[111])$ showing the range of shear modulus of electrolyte over which electrodeposition is stable and its dependence on the volume ratio $v$ of the cation and metal atom.}
\end{figure}

This interface has an anisotropic electrode on one side and isotropic solid electrolyte on the other. This is worth studying since the candidate material for anode, namely Li metal is highly anisotropic compared to other materials \cite{xuLi2017}. Fig. \ref{fig:chi} shows the value of the stability parameter $\chi$ for the three cases of \hkl[1 0 0], \hkl[1 1 0] and \hkl[1 1 1] crystallographic direction of Li perpendicular to the surface of solid electrolyte as a function of its shear modulus.  Investigation of the stability parameter for other facets gave the same general trend.
As observed in the isotropic-isotropic case, for $v<1$, $\chi$ increases with $G_s$ resulting in stability below the critical shear modulus value. For $v>1$, $\chi$ decreases with $G_s$, resulting in stability beyond a critical shear modulus. Each surface of Li has a different elastic response which results in different stresses at the interface. The stress results in different values of the stability parameter $\chi$ for the different surfaces. The stability diagram is then dependent on surface orientation of Li in contact with the solid electrolyte. Fig.~\ref{fig:stab-diag-anis} shows the stability diagram for different surfaces of Li metal anode in contact with a solid electrolyte. The nature of the stability diagram remains the same with two stable and two unstable regions. The stable regions are located below the critical curves for $v<1$ and above the critical curves for $v>1$.
However, the critical shear modulus curves shift depending on the specific surface of Li in contact with the solid electrolyte. In the $v>1$ region, for example, the surface \hkl(1 1 1) imposes stronger requirements on the shear modulus for stability than the \hkl(1 0 0) surface.  This presents opportunities for dendrite suppression if solid electrolytes could be made to preferentially comply along certain directions, for example, \hkl[1 1 1].

\subsection{Anisotropic-anisotropic interface}
The fully anisotropic interface equations were solved using the Stroh formalism with the stresses and deformations obtained using Eq. (\ref{eq:ustresultmat}). The orientation dependent elastic tensor of the two materials enters the problem through eigen vectors $\rtvec{a}$, $\rtvec{b}$ and solutions $p$ of the sextet Eq. (\ref{eq:degree6}). Due to the high dimensionality of the anisotropic problem (21 components of the elastic tensor, different surfaces in contact) and the absence of an analytical solution for the stability parameter, we assessed the stability of electrodeposition at anisotropic-anisotropic interfaces on a case by case basis instead of a high dimensional stability diagram. We considered several high Li-ion conducting solids including several obtained by Sendek ~\etal through large-scale screening of Li-containing compounds  \cite{sendek2017holistic} which have experimental \cite{crcelastic} or first-principles computed elastic tensor available \cite{Ahmad16uncertainty,Jain2013, de2015charting}. Overall, the solid electrolytes considered here include the major classes of thiophosphates (\ce{Li10GeP2S12}) \cite{Kamaya2011}, halides (\ce{LiI}) \cite{poulsen1981-halide}, garnets (\ce{Li5La3Ta2O12}) \cite{weppner2003-garnet}, phosphates (\ce{Li3PO4}) used in LiPON \cite{Yu1997-lipon, dudney2005-lipon}, sulfides (\ce{Li2S}) used in \ce{Li2S}-\ce{P2S5} solid electrolyte \cite{tatsumisago2013-sulfide} and alloys (\ce{LiCu3}). Low index surfaces of Li and solid electrolyte were considered at the interface.

Table \ref{tab:chianiso} lists the values of the stability parameter $\chi$ for different interfaces between Li metal anode and a solid electrolyte. The crystallographic directions along $\rtvec{e_1}$ and $\rtvec{e_2}$ are $\rtvec{u_2}$ and $\rtvec{v_2}$ for the two materials in contact. The molar volume ratio $v$ was calculated using coordination-dependent values of ionic radii tabulated  by Shannon \cite{shannon1976revised} and Marcus ~\etal \cite{marcus-ionvolumes} (See also Ref. \onlinecite{ahmad2017stability}). The role of anisotropy is evident from the drastic changes in the stability parameter upon changing the interface crystallographic directions. Inorganic solid electrolytes with a lower shear modulus generally have lower stability parameters as should be expected from the isotropic case for $v<1$. Unfortunately, none of the solid electrolytes we investigated have a negative stability parameter i.e. an interface stable against growth of dendrites.

\begin{table}[htbp]
\caption{\label{tab:chianiso}%
Stability parameter of fully anisotropic Li-solid electrolyte interfaces. The crystallographic orientation of the solids can be identified from crystallographic directions $\rtvec{u_2}$ and $\rtvec{v_2}$ which are lie along $\rtvec{e_1}$ and $\rtvec{e_2}$ respectively.}
\begin{ruledtabular}
\begin{tabular}{lccccc}
Electrolyte & \multicolumn{2}{c}{Electrode} & \multicolumn{2}{c}{Electrolyte} & $\chi$\\
material & $\rtvec{v_2}$ & $\rtvec{u_2}$ & $\rtvec{v_2}$ & $\rtvec{u_2}$ & (kJ/mol$\cdot$nm)\\
\colrule
\begin{tabular}{l}
\ce{Li10GeP2S12}\\
%(Thiophosphate)\\
$v=0.151$
\end{tabular} & 
\begin{tabular}{c}\\
\hkl[0 1 0]\\
\hkl[1 1 0]\\
\hkl[0 1 1]\\
\hkl[1 1 1]\\
\hkl[0 1 0]\\
\hkl[0 1 0]\\
\hkl[1 1 0]\\
\hkl[0 1 0]\\
\hkl[1 1 0]\\
\end{tabular} & 
\begin{tabular}{c}\\
\hkl[1 0 0]\\
\hkl[1 -1 0]\\
\hkl[1 0 0]\\
\hkl[79 -58 -21]\\
\hkl[1 0 0]\\
\hkl[1 0 0]\\
\hkl[1 -1 0]\\
\hkl[1 0 0]\\
\hkl[1 -1 0]
\end{tabular} &
\begin{tabular}{c}\\
\hkl[0 1 0]\\
\hkl[0 1 0]\\
\hkl[0 1 0]\\
\hkl[0 1 0]\\
\hkl[1 1 0]\\
\hkl[1 1 1]\\
\hkl[1 1 0]\\
\hkl[0 0 1]\\
\hkl[0 0 1]\\
\end{tabular} &
\begin{tabular}{c}\\
\hkl[1 0 0]\\
\hkl[1 0 0]\\
\hkl[1 0 0]\\
\hkl[1 0 0]\\
\hkl[1 -1 0]\\
\hkl[95  -56  -19]\\
\hkl[1 -1 0]\\
\hkl[1 0 0]\\
\hkl[1 0 0]
\end{tabular} &
\begin{tabular}{d}\\
7524.5\\
10990.3\\
7781.1\\
9161.7\\
7609.6\\
8348.0\\
11075.4\\
10234.9\\
14898.8
\end{tabular} 
\\
\begin{tabular}{l}
\ce{LiI}\\
%(Halide)\\
$v=0.099$
\end{tabular} & 
\begin{tabular}{c}\\
\hkl[0 1 0]\\
\hkl[1 1 0]\\
\hkl[1 1 1]\\
\hkl[0 1 0]\\
\hkl[0 1 0 ]\\
\hkl[0 1 0] \\
\hkl[1 1 0]\\
\hkl[1 1 0]
\end{tabular} & 
\begin{tabular}{c}\\
\hkl[1 0 0]\\
\hkl[1 -1 0]\\
\hkl[79 -58 -21]\\
\hkl[1 0 0]\\
\hkl[1 0 0]\\
\hkl[1 0 0]\\
\hkl[1 -1 0]\\
\hkl[1 -1 0]\\
\end{tabular} &
\begin{tabular}{c}\\
\hkl[0 1 0]\\
\hkl[0 1 0]\\
\hkl[0 1 0]\\
\hkl[1 1 0]\\
\hkl[1 1 1]\\
\hkl[0 1 1]\\
\hkl[1 1 0]\\
\hkl[1 1 1]
\end{tabular} &
\begin{tabular}{c}\\
\hkl[1 0 0]\\
\hkl[1 0 0]\\
\hkl[1 0 0]\\
\hkl[1 -1 0]\\
\hkl[79 -58 -21]\\
\hkl[1 0 0]\\
\hkl[1 -1 0]\\
\hkl[79 -58 -21]
\end{tabular} &
\begin{tabular}{d}\\
6525.4\\
10776.0\\
8526.9\\
6619.8\\
7530.8\\
7903.4\\
10870.5\\
11950.4
\end{tabular} 

\\
\begin{tabular}{l}
\ce{Li5La3Ta2O12}\\
%(Garnet)\\
$v=0.085$
\end{tabular} & 
\begin{tabular}{c}\\
\hkl[0 1 0]\\
\hkl[1 1 0]\\
\hkl[1 1 1]\\
\hkl[0 1 0]\\
\hkl[0 1 0]\\
\hkl[0 1 0]\\
\hkl[1 1 0]\\
\hkl[1 1 0]
\end{tabular} & 
\begin{tabular}{c}\\
\hkl[1 0 0]\\
\hkl[1 -1 0]\\
\hkl[79 -58 -21]\\
\hkl[1 0 0]\\
\hkl[1 0 0]\\
\hkl[1 0 0]\\
\hkl[1 -1 0]\\
\hkl[1 -1 0]
\end{tabular} & 
\begin{tabular}{c}\\
\hkl[0 1 0]\\
\hkl[0 1 0]\\
\hkl[0 1 0]\\
\hkl[1 1 0]\\
\hkl[1 1 1]\\
\hkl[0 1 1]\\
\hkl[1 1 0]\\
\hkl[1 1 1]
\end{tabular} & 
\begin{tabular}{c}\\
\hkl[1 0 0]\\
\hkl[1 0 0]\\
\hkl[1 0 0]\\
\hkl[1 -1 0]\\
\hkl[79 -58 -21]\\
\hkl[1  0 0]\\
\hkl[1 -1 0]\\
\hkl[79 -58 -21]
\end{tabular} & 
\begin{tabular}{d}\\
44897.9\\
50014.9\\
47980.7\\
44924.5\\
46113.1\\
46583.4\\
50041.5\\
51257.6
\end{tabular} 
\\
\begin{tabular}{l}
\ce{Li3PO4}\\
%(Phosphate)\\
$v=0.098$
\end{tabular} & 
\begin{tabular}{c}\\
\hkl[0 1 0]\\
\hkl[1 1 0]\\
\hkl[0 1 1]\\
\hkl[1 1 1]\\
\hkl[0 1 0]\\
\hkl[0 1 0]\\
\hkl[0 1 0]\\
\hkl[0 1 0]
\end{tabular} & 
\begin{tabular}{c}\\
\hkl[1 0 0]\\
\hkl[1 -1 0]\\
\hkl[1 0 0]\\
\hkl[79 -58 -21] \\
\hkl[1 0 0]\\
\hkl[1 0 0]\\
\hkl[1 0 0]\\
\hkl[1 0 0]
\end{tabular} &
\begin{tabular}{c}\\
\hkl[0 1 0]\\
\hkl[0 1 0]\\
\hkl[0 1 0]\\
\hkl[0 1 0]\\
\hkl[1 1 0]\\
\hkl[1 1 1]\\ 
\hkl[0 1 1]\\
\hkl[1 0 0]\\ 
\end{tabular} &
\begin{tabular}{c}\\
\hkl[1 0 0]\\
\hkl[1 0 0]\\
\hkl[1 0 0]\\
\hkl[1 0 0]\\
\hkl[78 -51 0]\\
\hkl[90 -31 -10 ]\\
\hkl[1 0 0]\\
\hkl[0 1 0]
\end{tabular} &
\begin{tabular}{d}\\
35942.7\\
41575.6\\
36586.7\\
39221.0\\
35466.1\\
35136.1\\
35671.5\\
34696.9
\end{tabular} 
\\
\begin{tabular}{l}
\ce{Li2S}\\
%(Sulfide)\\
$v=0.066$
\end{tabular} & 
\begin{tabular}{c}\\
\hkl[0 1 0]\\
\hkl[1 1 0]\\
\hkl[0 1 1]\\
\hkl[1 0 1]\\
\hkl[1 1 1]\\
\hkl[0 1 0]\\
\hkl[0 1 0]\\
\hkl[0 1 0]
\end{tabular} & 
\begin{tabular}{c}\\
\hkl[1 0 0]\\
\hkl[1 -1 0]\\
\hkl[1 0 0]\\
\hkl[50 -71 -50]\\
\hkl[79 -58 -21]\\
\hkl[1 0 0]\\
\hkl[1 0 0]\\
\hkl[1 0 0]
\end{tabular} & 
\begin{tabular}{c}\\
\hkl[0 1 0]\\
\hkl[1 1 0]\\
\hkl[0 1 1]\\
\hkl[1 0 1]\\
\hkl[1 1 1]\\
\hkl[0 1 0]\\
\hkl[0 1 0]\\
\hkl[0 1 0]
\end{tabular} & 
\begin{tabular}{c}\\
\hkl[0 1 0]\\
\hkl[1 1 0]\\
\hkl[0 1 1]\\
\hkl[1 0 1]\\
\hkl[1 1 1]\\
\hkl[0 1 0]\\
\hkl[0 1 0]\\
\hkl[0 1 0]
\end{tabular} & 
\begin{tabular}{d}\\
26619.8\\
32705.2\\
27536.2\\
31769.4\\
30338.8\\
26637.4\\
27666.1\\
27216.4
\end{tabular} \\

\begin{tabular}{l}
\ce{LiCu3}\\
%(Alloy)\\
$v=0.738$
\end{tabular} & 
\begin{tabular}{c}\\
\hkl[0 1 0]\\
\hkl[1 1 0]\\
\hkl[0 1 1]\\
\hkl[1 1 1]\\
\hkl[0 1 0]\\
\hkl[0 1 0]\\
\hkl[1 1 0]\\
\hkl[0 1 0]\\
\hkl[1 1 0]\\
\end{tabular} & 
\begin{tabular}{c}\\
\hkl[1 0 0]\\
\hkl[1 -1 0]\\
\hkl[1 0 0]\\
\hkl[79 -58 -21]\\
\hkl[1 0 0]\\
\hkl[1 0 0]\\
\hkl[1 -1 0]\\
\hkl[1 0 0]\\
\hkl[1 -1 0]
\end{tabular} &
\begin{tabular}{c}\\
\hkl[0 1 0]\\
\hkl[0 1 0]\\
\hkl[0 1 0]\\
\hkl[0 1 0]\\
\hkl[1 1 0]\\
\hkl[1 1 1]\\
\hkl[1 1 0]\\
\hkl[0 0 1]\\
\hkl[0 0 1]\\
\end{tabular} &
\begin{tabular}{c}\\
\hkl[1 0 0]\\
\hkl[1 0 0]\\
\hkl[1 0 0]\\
\hkl[1 0 0]\\
\hkl[1 -1 0]\\
\hkl[95  -56  -19]\\
\hkl[1 -1 0]\\
\hkl[1 0 0]\\
\hkl[1 0 0]
\end{tabular} &
\begin{tabular}{d}\\
8005.3\\
11995.8\\
10457.7\\
11628.6\\
9910.0\\
14296.6\\
13900.5\\
11148.1\\
15682.2
\end{tabular} 

\end{tabular}
\end{ruledtabular}

\end{table}

The volume ratio $v$ here deserves some discussion. Inorganic solid electrolytes generally have $0<v<1$. Li alloys have $v$ close to 1 while compounds with low Li coordination number have lower $v$. If we observe the overall range of $\chi$ for different materials while varying surface orienations, \ce{Li10GeP2S12} and \ce{LiI} with low shear modulus have stability parameter closer to zero than the other high shear modulus compounds, while alloy \ce{LiCu3} with high volume ratio $v$ has $\chi$ closer to zero despite a high shear modulus (36 GPa). On comparing $\chi$ for different surface orientations for a given material, we observe that the orientation with $\rtvec{v_2}=\mathscr{V}(\hkl[0 1 0])$ and $\rtvec{u_2}=\mathscr{V}(\hkl[1 0 0])$  for  both the electrode and electrolyte (first entry in the table for each material) has the lowest stability parameter. This is, thus, the most compliant arrangement. These trends could be used to search for solid electrolytes for stable electrodeposition.
%\vv{Can you include a small discussion of the table.  Stable which are the "soft" solid electrolytes.  Discuss a bit about the range of v for different inorganic materials.   Metal alloys ..  Give some general material design guideline --  (111) compliance.}

\section{Discussion}\label{sec:disc}
We discuss some general principles which can be used to make sense of the stability diagrams. For the isotropic case, the deviatoric term is zero at $G_s=0$ which means the sign of hydrostatic term determines the stability in the zero shear modulus limit. At non-zero shear modulus, this term is always found to be destabilizing\footnote{This can be seen easily for the isotropic-isotropic interface using the analytical expression for $\chi$ given in Supplemental Material of Ref. \onlinecite{ahmad2017stability}. For the case of isotropic electrolyte in contact with anisotropic electrode, we found this to be true for all facets of Li investigated.}. The existence of stability regions for isotropic solid electrolyte case follows from the dependence of stability parameter $\chi$ on the hydrostatic term alone. $\chi$ gives the electrochemical potential change of the electron at a peak in the interface ($\Delta \mu_{e^-}=\chi$ when $\cos(kx_1)=1$). For $v<1$, the hydrostatic term in Eq. (\ref{eq:mu}) is stabilizing when $\Delta p_e + \Delta p_s$ is negative. At the peak (e.g. $x_1=0$), tensile stress is generated at the electrode side of the interface and compressive at the electrolyte side, resulting in $\Delta p_e<0$ and $\Delta p_s>0$. When $G_s$ is low, $|\Delta p_s|\ll |\Delta p_e|$ which will make this term stabilizing. For $v>1$, we require $\Delta p_e + \Delta p_s>0$ for stability and this will occur at high $G_s$ with $|\Delta p_s|\gg |\Delta p_e|$. This argument explains the stable regimes at the bottom left and top right in Fig. \ref{fig:stab-diag-anis}. The two unstable regions in the isotropic-isotropic and isotropic-anisotropic cases are separated by the black line and are different phases since one cannot go from one phase to another without passing through $\chi=0$.

\section{Conclusions}\label{sec:concl}
We used the Stroh formalism to analyze the stability of electrodeposition at different kinds of solid-solid interfaces. The results obtained for the isotropic case using this formalism match the results of previous studies. The isotropic solid electrolyte in contact with the anisotropic Li metal anode has a qualitatively similar stability diagram with the critical shear modulus curves depending on the crystallographic orientation of Li at the interface. Both the isotropic-isotropic and isotropic-anisotropic interface stability diagrams point towards development of low shear modulus inorganic solid state electrolytes (which generally have $0<v<1$) for obtaining stable electrodeposition. For completely anisotropic interfaces, we find that the stability parameter is highly dependent on the crystallographic orientation of the solids in contact.  In the context of Li metal anodes in contact with electrolyte having $v>1$, compliance along \hkl[1 0 0] direction for a solid electrolyte leads to less stringent requirements on the modulus of the solid electrolyte while \hkl[1 1 1] leads to more stringent requirements. A similar analysis might also be useful in problems on stability of solid-solid interfaces encountered in other areas, for example, epitaxial thin films when the materials have a high degree of anisotropy.

\begin{acknowledgments}
We thank S. Hulikal for helpful discussions. Z. A. acknowledges support from the Advanced Research Projects Agency-Energy Integration and Optimization of Novel Ion Conducting Solids (IONICS) program under Grant No. \texttt{DE-AR0000774}. V. V. gratefully acknowledges support from the U.S. Department of Energy, Energy Efficiency and Renewable Energy Vehicle Technologies Office under Award No. \texttt{DE-EE0007810}.
\end{acknowledgments}

\appendix

\section{Stroh Formalism for degenerate case of isotropic material}\label{app:degen}
For isotropic materials with shear modulus $G$ and Poisson's ratio $\nu$, all three solutions $p_\alpha$ of Eq. (\ref{eq:degree6}) are equal to i ($\text{i}^2=-1$). The matrices $\rtten{F}$, $\rtten{A}$ and $\rtten{B}$ used to determine the solution are:

\begin{eqnarray*}
\rtten{F}=\begin{bmatrix}f(x_1+p_1x_2) & x_2 f'(x_1+p_1x_2) & 0\\
0 & f(x_1+p_2x_2) & 0\\
0 & 0 & f(x_1+p_3x_2)
\end{bmatrix}
\end{eqnarray*}

\begin{eqnarray*}
\rtten{A}=\psi \begin{bmatrix}1 & -\text{i}\gamma & 0\\
\text{i} & -\gamma & 0\\
0 & 0 & \varepsilon
\end{bmatrix};
\rtten{B}=\begin{bmatrix}2\text{i} & 1 & 0\\
-2 & -\text{i} & 0\\
0 & 0 & \text{i}\varepsilon
\end{bmatrix}\\
\psi=\frac{1}{\sqrt{8G(1-\nu)}}, \gamma=\frac{1}{2}(3-4\nu), \varepsilon= (1-\text{i})\sqrt{2(1-\nu)}
\end{eqnarray*}

Eq. (\ref{eq:ustrresult}) can then be used to obtain deformation and stress fields for isotropic materials.

\section{Transformation of elastic tensor in Voigt form}\label{app:ctrans}
Let $\rtten{C}$ be the $6\times 6$ elastic tensor in Voigt form associated with a particular coordinate system and $\rtten{\tilde{C}}$ be the transformed elastic tensor under rotation $\rtten{Q}$. Then $\rtten{\tilde{C}}$ can be calculated as \cite{ting-anis, bower2009mechanics}:
\begin{eqnarray}
\rtten{\tilde{C}}=\rtten{KCK^T}
\end{eqnarray}
where $\rtten{K}$ is $6\times 6$ tensor given by:
\iffalse
\begin{gather*}
\rtten{K}=\begin{bmatrix} \rtten{K_1} & \rtten{2K_2}\\
\rtten{K_3} & \rtten{K_4}
\end{bmatrix}\\
\rtten{K_1}=\begin{bmatrix}Q_{11}^2 & Q_{12}^2 & Q_{13}^2\\
Q_{21}^2 &  Q_{22}^2 & Q_{23}^2\\
Q_{31}^2 & Q_{32}^2 & Q_{33}^2
\end{bmatrix}\\
\rtten{K_2}=\begin{bmatrix} Q_{12}Q_{13} & Q_{13}Q_{11} & Q_{11}Q_{12}\\
Q_{22}Q_{23} & Q_{23}Q_{21} & Q_{21}Q_{22}\\
Q_{32}Q_{33} & Q_{33}Q_{31} & Q_{31}Q_{32}
\end{bmatrix}\\
\rtten{K_3}=\begin{bmatrix}Q_{21}Q_{31}& Q_{22}Q_{32}& Q_{23}Q_{33}\\
Q_{31}Q_{11} & Q_{32}Q_{12} & Q_{33}Q_{13}\\
Q_{11}Q_{21} & Q_{12}Q_{22}& Q_{13}Q_{23}
\end{bmatrix}\\
[\rtten{K_4}]_{ij}=Q_{\text{mod}(i+1,3)\text{mod}(j+1,3)}Q_{\text{mod}(i+2,3)\text{mod}(j+2,3)}\\
+Q_{\text{mod}(i+1,3)\text{mod}(j+2,3)}Q_{\text{mod}(i+2,3)\text{mod}(j+1,3)}\\
%\rtten{K_4}=\begin{bmatrix}Q_{22}Q_{33}+Q_{23}%Q_{32} & Q_{23}Q_{31}+Q_{21}Q_{33}& Q_{21}%Q_{32}+Q_{22}Q_{31}\\
%Q_{32}Q_{13}+Q_{33}Q_{12}& Q_{33}Q_{11}%+Q_{31}Q_{13}& Q_{31}Q_{12}+Q_{32}Q_{11}\\
%Q_{12}Q_{23}+Q_{13}Q_{22}& Q_{13}Q_{21}%+Q_{11}Q_{23}& Q_{11}Q_{22}+Q_{12}Q_{21}
%\end{bmatrix}
\end{gather*}
\fi
\begin{gather*}
\rtten{K}=\begin{bmatrix} \rtten{K_1} & \rtten{2K_2}\\
\rtten{K_3} & \rtten{K_4}
\end{bmatrix}\\
[\rtten{K_1}]_{ij}=[\rtten{Q}]_{ij}^2\\
[\rtten{K_2}]_{ij}=[\rtten{Q}]_{i\text{mod}(j+1,3)}[\rtten{Q}]_{i\text{mod}(j+2,3)}\\
[\rtten{K_3}]_{ij}=[\rtten{Q}]_{\text{mod}(i+1,3)j}[\rtten{Q}]_{\text{mod}(i+2,3)j}\\
[\rtten{K_4}]_{ij}=[\rtten{Q}]_{\text{mod}(i+1,3)\text{mod}(j+1,3)}[\rtten{Q}]_{\text{mod}(i+2,3)\text{mod}(j+2,3)}\\
+[\rtten{Q}]_{\text{mod}(i+1,3)\text{mod}(j+2,3)}[\rtten{Q}]_{\text{mod}(i+2,3)\text{mod}(j+1,3)}\\
\end{gather*}
where 
\begin{equation*}
\text{mod}(i,3)=\begin{cases} 
i & i\leq 3\\
i-3 & i>3
\end{cases}
\end{equation*}

Next, we show how to determine the rotation matrix $\rtten{Q}$ and $\rtvec{u_2}$ given $\rtvec{v_2}$.

\emph{Example for a cubic crystal.}-
Let $\rtvec{v_2}$ coincide with [110] direction of the crystal or $\rtvec{v_2}=\mathscr{V}([110])$. Then the rotation matrix $\rtten{Q}$ obtained by following the procedure mentioned in Sec. \ref{sec:ctrans} is given by:
\begin{equation*}
\rtten{Q}=\begin{bmatrix}\frac{1}{\sqrt{2}} & -\frac{1}{\sqrt{2}} & 0\\
\frac{1}{\sqrt{2}} & \frac{1}{\sqrt{2}} & 0\\
0 & 0 & 1
\end{bmatrix}
\end{equation*}
The crystallographic direction which aligns along $\rtvec{e_1}$ due to this rotation is given by:
\begin{eqnarray*}
\rtvec{u_2}=\rtten{Q}^{-1}\mathscr{V}\left(\begin{bmatrix}1\\
0\\
0
\end{bmatrix}\right)=a\begin{bmatrix}\frac{1}{\sqrt{2}}\\
-\frac{1}{\sqrt{2}}\\
0
\end{bmatrix}=\mathscr{V}\left( \hkl[1 -1 0]\right)
\end{eqnarray*}
Hence, $\rtvec{u_2}$ corresponds to the $\hkl[1 -1 0]$ direction of the crystal. Similarly, the other combinations ($\rtvec{v_2}$,$\rtvec{u_2}$) along ($\rtvec{e_2}$,$\rtvec{e_1}$) are ($\hkl[0 1 1]$,$\hkl[1 0 0]$),  ($\hkl[2 2 1]$,$\hkl[11 -10  -2]$). Note that $\rtvec{u_2}$ is always perpendicular to $\rtvec{v_2}$. For non-cubic crystals, care must be taken to differentiate the crystallographic axes (in Miller index notation) from the actual direction vectors for calculating $\rtten{Q}$ and $\rtvec{u_2}$.
\bibliography{../../../solid-electrolytes/bibtexs/refs}% Produces the bibliography via BibTeX.

\end{document}